\documentclass{elsart41}
\usepackage{graphics}
\usepackage{graphicx}
\usepackage{amssymb}
\begin{document}

\begin{frontmatter}



\title{Theory of Fano-Kondo effect in quantum dot systems:
temperature dependence of the Fano line shapes}

%

\author[AA]{I. Maruyama\corauthref{Maruyama}},
\ead{maru@th.physik.uni-bonn.de}
\author[BB]{N. Shibata},\,
\author[CC]{K. Ueda}

\address[AA]{Physikalisches Institut der Universitaet Bonn, Nussallee 
12, 53115 Bonn, Germany}
\address[BB]{Dept. of Basic Science, Univ. of Tokyo
3-8-1 Komaba, Meguro, Tokyo 153-8902, Japan}
\address[CC]{Institute for Solid State Physics, Univ. of Tokyo,
5-1-5 Kashiwa-no-ha, Kashiwa, Chiba 277-8581, Japan}

\corauth[Maruyama]{Corresponding author. Tel: +49(228)73-3719,
fax: +49(228)73-9336}

\begin{abstract}
The Fano-Kondo effect in zero-bias conductance is studied based on a
theoretical model for the T-shaped quantum dot by the finite
temperature density matrix renormalization group method.  The
modification of the two Fano line shapes at much higher temperatures
than the Kondo temperature is also investigated by the effective Fano
parameter estimated as a fitting parameter.
\end{abstract}

\end{frontmatter}

Conductance through quantum dots (QDs) as a function of gate voltage
generally shows asymmetric peak structures due to the Fano effect.
This structure is characterized by the Fano asymmetric parameter $q$.
Concerning the theories of the T-shaped QD under the zero-bias
condition, the Fano-Kondo effect studied so far has been limited to
the $q=0$ case\cite{KangCKS01,TorioHCP02,AligiaP02}, where conductance
has dip structures at high temperatures and shows the anti-Kondo
resonance at low temperatures.

To consider the finite-$q$ case of the T-shaped QD, we introduce a
simple tight-binding model\cite{MaruyamaSU04-2} which is a
one-dimensional Anderson model with an additional gate voltage
$\epsilon_0$ applied to the 0-th site.
\begin{eqnarray}
H&=& - \sum_{i\sigma} t_{i,i+1} ( c^\dagger_{i\sigma} c_{i+1\sigma} + 
\mbox{h.c.} ) \nonumber \\ && +\epsilon_0 \sum_{\sigma} 
c^\dagger_{{0}\sigma} c_{{0}\sigma}
-v_d \sum_{\sigma} ( d^\dagger_{\sigma} c_{0\sigma} + \mbox{h.c.} ) 
\nonumber \\ && +\epsilon_d \sum_{\sigma} d^\dagger_{\sigma} d_{\sigma} 
+U_d d^\dagger_{\uparrow} d_{\uparrow} d^\dagger_{\downarrow} d_{\downarrow},
\label{eq:H}
\\
t_{i,i+1}&:=&
\left\{
\begin{array}{cc}
v_0, & \mbox{when $i=0$ or $-1$,} \\
t = 1, & \mbox{others,}\\
\end{array}
\right.
\end{eqnarray}
where QD has a single level, $\epsilon_d$, and an on-site Coulomb
interaction, $U_d$.  It is important for this model that the Fano
parameter is changed by the additional gate voltage:
$q=\epsilon_0/\Delta_0=\epsilon_0 t/ 2v_0^2$.  Actually, zero-bias
conductance $g$ normalized by $2e^2/h$ is given by this parameter $q$
and the local Green's function at the QD, $G_d (\omega)$.  The local
Green's function can be obtained by the finite temperature density
matrix renormalization group (F$T$-DMRG) method after a numerical
analytic continuation either by the maximum entropy method (MEM) or
the Pad\'e approximation.  The results are characterized by two Fano
asymmetric peaks at high temperatures and by the Fano-Kondo plateau
inside a Fano peak at low temperatures.  For further details, see
ref.~\cite{MaruyamaSU04-2}.

Strictly speaking, the Kondo temperature depends on the gate voltage,
which can be written around the symmetric case: $T_{\rm
K}(\epsilon_d)\simeq \sqrt{\Delta_d U_d/2}\exp (\pi \epsilon_d
(\epsilon_d + U_d)/ (2\Delta_d U_d) ) $, where $\Delta_d = v_d^2 /
\Delta_0 (1 + q^2)$.  When the QD is doubly occupied and empty, the
lowest characteristic temperature is the minimum of $U_d$ and $\pi
\Delta_d$.  This characteristic temperature is smoothly connected to
the Kondo temperature around the symmetric case\cite{HewsonOM04}.
Especially, two Fano peaks are modified at temperatures $T <
\mbox{min}(U_d,\pi \Delta_d)$, which is much larger than $T_{\rm
K}(-U_d/2)$.

\begin{figure}[!ht]
\begin{center}
\includegraphics[width=0.45\textwidth]{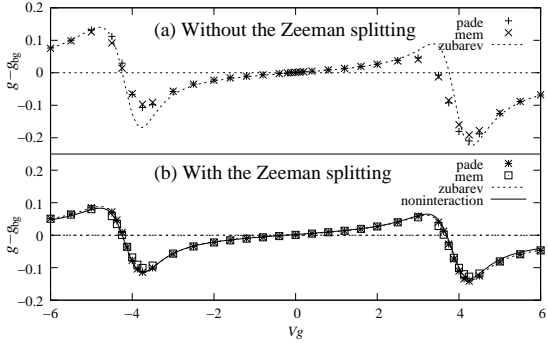}
\end{center}
\caption{(a) Conductance in a well separated case: $T/t = 0.2, q =
-0.8, \Delta_0/t = 1, \Delta_d/t = 0.15 (v_d = 0.5)$ and $U_d/t = 8$.
$g_{\rm bg}$ is a background current.  (b) Conductance under a large
Zeeman splitting $\Delta \epsilon_d = 4$ for $U_d=4$.  Other
parameters are the same as those of (a).  The line and the points are
very close to the one of the non-interacting results with $\Delta
\epsilon_d=8, U_d=0$.  }
\label{fig1}
\end{figure}

In this paper, we study the modification of the Fano shape at the
intermediate temperatures.  We also compare the numerical results with
the conductance obtained by the Zubarev's approximation, which is
valid at high temperatures.  One can see the difference between the
numerical results and the Zubarev's approximation in
Fig.~\ref{fig1}(a).  The difference becomes smaller when a local
magnetic field is applied on QD to suppress the Kondo effect as shown
in Fig.~\ref{fig1}(b).  Moreover, these results are very close to the
non-interacting results.  This fact means that the ``one body level''
picture is good if large Zeeman splitting is applied.  In other words,
the left (right) part of the right (left) peak in Fig.~\ref{fig1}(a)
is modified due to the many body effects which are described by beyond
the Zubarev's approximation.

To determine the modification of two Fano line-shapes quantitatively,
we estimate the effective Fano asymmetric parameter $q_T$ at a given
temperature as a fitted parameter.  In order to fit the shape of the
Fano peaks, there are several possible ways as used in experimental
analyses \cite{ZachariaGGKK01}.  The simplest way may be to fit with
the function,
\begin{eqnarray}
g(V_{g})&=& g_{bg} + g_{a} { ( e(V_{g}) + q_T )^2 \over e(V_{g})^2 +1 },
\\
e(V_{g})&=& {V_{g} - \Delta V_{g}\over \Delta_{dT} },
\label{eq:FanoFit}
\end{eqnarray}
with five fitting parameters $g_{bg}, g_{a}, q_T, \Delta V_{g}$ and
$\Delta_{dT}$.  In order to fit two Fano structures, we separate them
into two ranges: $-6<V_g<-2$ and $2<V_g<6$ and fit the data in each
range with the five parameters.

\begin{figure}[!ht]
\begin{center}
\includegraphics[width=0.45\textwidth]{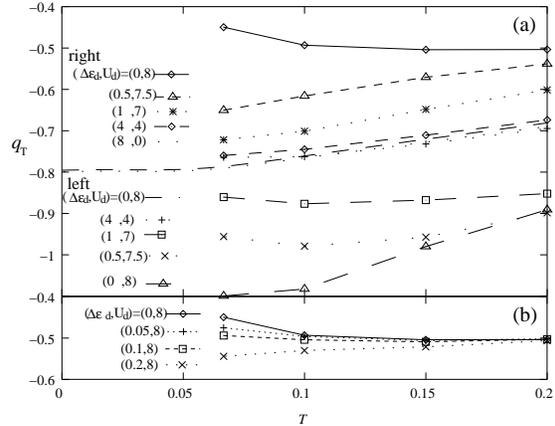}
\end{center}
\caption{Fitted results of $q_T$ as a function of temperature.  The
fixed range, $-6 \leq V_g\leq -2$, is used for the fitting for the
left Fano structure ($2\leq V_g\leq 6$ for the right).  The data are
obtained for the results of $q=-0.8, \Delta_0=1, \Delta_d=0.15$ with
varying $\Delta \epsilon_d$ and $U_d$.
\label{fig:qT}}
\end{figure}

We plot the temperature dependence of $q_T$ in Fig.\ref{fig:qT} with
several lines by changing the Zeeman splitting energy $\Delta
\epsilon_d$.  The difference of $q_T$ between left and right Fano
structures disappears when strong Zeeman splitting is applied.
Especially, the difference between the data in $\Delta \epsilon_d=0$
case becomes larger with decreasing temperature.  Finally at zero
temperature, the asymptotic behavior of $q_T$ is expected to go toward
$0$ and $-\infty$ because the Fano asymmetric peaks at the
high-temperature regime become a peak ($|q_T|\rightarrow \infty$) or a
dip ($q_T\rightarrow 0$) in the low-temperature regime.  In this
sense, the temperature dependence of the estimated Fano parameter
$q_T$ can be regarded as a precursor of the Fano-Kondo effect for a
finite $q$ case.  From the realistic viewpoint, it is important that
this many-body effect is observable at relatively high temperatures
$T/\pi \Delta_d \sim 1$ which is larger than the lowest Kondo
temperature at the plateau.


\end{document}